\documentclass[aps,prl,showpacs,twocolumn,floatfix,reprint]{revtex4-1}
\usepackage{amsfonts}
\usepackage{mathrsfs}
\usepackage{color}
\usepackage{xspace}
\usepackage{epstopdf}
\usepackage{subfigure}
\usepackage{longtable}
\usepackage{xspace}
\usepackage{multirow}
\usepackage{tabularx}
\usepackage{amsmath}
\usepackage{amssymb}
\usepackage{graphicx}
\usepackage{dcolumn}
\usepackage{bm}
\usepackage{hyperref}

\begin{document}

\title{Majorana fermions in quasi-1D and higher dimensional ultracold optical
lattices}
\author{Chunlei Qu$^{1}$}
\author{Ming Gong$^{2}$}
\author{Yong Xu$^{1}$}
\author{Sumanta Tewari$^{3}$}
\author{Chuanwei Zhang$^{1}$}
\thanks{Email: chuanwei.zhang@utdallas.edu}

\begin{abstract}
We show that Majorana fermions (MFs) exist in two- and three-dimensional
(2D,3D) fermionic optical lattices with strictly 1D spin-orbit coupling (SOC)
which has already been realized in experiments. For a quasi-1D topological
BCS superfluid, there are multiple MFs at each end which are topologically
protected by a chiral symmetry. In the generalization to higher dimensions,
the multiple MFs form a zero energy flat band. An additional experimentally
tunable in-plane Zeeman field drives the system to a topological Fulde-Ferrell
(FF) superfluid phase. We find that even though the multiple MFs are robust
against the in-plane Zeeman field if the order parameters at the different
chains are enforced to be identical, they are destroyed in the
self-consistently obtained FF phase where the order parameters are
inhomogeneous on the boundaries. Our results are useful to guide the
experimentalists on searching for MFs in the context of ultracold fermionic
atoms.
\end{abstract}

\affiliation{$^{1}$Department of Physics, the University of Texas at Dallas,
Richardson, TX 75080, USA \\
$^{2}$Department of Physics and Centre for Quantum Coherence, The Chinese University of Hong Kong, Shatin, N.T.,
Hong Kong, China \\
$^{3}$Department of Physics and Astronomy, Clemson University, Clemson, South
Carolina 29634, USA}
\date{\today}
\maketitle


{\indent{\em Introduction.}}--- MFs, quantum particles which are their own
anti-particles, have attracted a lot of attention because of their topological
properties and the potential applications in fault-tolerant topological
quantum computation~\cite{RMP1,RMP2,RMP3}. Many solid state materials have
been predicted to be candidates for the realization of MFs~\cite{Kitaev,
Read-Green,Tewari-Strontium,Fu2008,Sau2010,Tewari-Annals,Alicea2010,Long-PRB,
Roman,Oreg2010,Mao,Williams,KTLaw}. Even though experimental progress in the
solid state systems has been made in the past few years and possible signatures
of MFs have been observed~\cite{Mourik,Deng,Das,Rokhinson,Veldhorst,Churchill,
Finck,Appelbaum,Stanescu}, a ``smoking gun" signature of MFs is still lacking
due to many factors influencing the measurement results in solid state
materials~\cite{Kells,Liu, Stanescu-Tewari-2,Dassarma,Roy-Tewari}. On the other
hand, ultracold atoms provide an ideal playground for the quantum simulations
of many condensed matter systems because they are clean and highly controllable
in the system parameters. The recent realization of SOC in BEC~\cite{SOBEC1,
SOBEC2,SOBEC3,SOBEC4} and Fermi gases~\cite{ SOFermi1,SOFermi2} paves a way for
the observation of MFS in cold atoms~\cite{Zhang2008,Sato2009,GM,SLZhu,Carlos}.
In this context, many schemes for the creation and observation of MFs in a 1D
cold atom quantum wire have been studied~\cite{Jiang-2011,Wei-2012,Liu-2012,LiuXJ}.

The realistic experiments in ultracold atoms are not on strictly 1D systems,
which motivates our present study on the existence and properties of MFs in
higher dimensional ultracold atom systems~\cite{Sato-2013}. The necessity of studying the
physics beyond 1D systems also arises from the failure of mean field theory
in 1D where there is no long range ordering due to Mermin-Wagner
theorem. The inclusion of a weak tunneling in the transverse directions in a
quasi-1D system could effectively suppress the quantum fluctuations and
stabilize the mean field superfluid order~\cite{Liao}. However, the presence
of such transverse tunneling terms, even if treated as a perturbation, may
pairwise couple the MFs and create a gap in the low energy spectrum. It follows
that, unless the number of chains in the transverse directions is odd (which is
difficult to control experimentally), the system of coupled chains may not
support any MFs at all. Because if this, whether or not MFs exist in weakly coupled
quasi-1D (with finite number of chains), 2D, and 3D cold atom systems with
artificial SOC and Zeeman fields has remained an important open question both
theoretically and experimentally.

In this paper we show that multiple localized zero energy MFs still exist in a
multi-chain system in a wide range of parameter space. In the limit of infinite
number of chains in the transverse directions, the MFs form a zero energy flat
band that can be probed experimentally. Our work is based on a chiral symmetric
analysis of a multi-chain system which is not applicable when there is a nonzero
in-plane Zeeman field~\cite{Sumanta1,Sumanta2,Diez-2012,BDI}. Thus, in addition
to the existence of the MFs we also explore their topological robustness against
an additional (in-plane) Zeeman field which may give rise to spatially inhomogeneous
order parameters in the resultant FF phase~\cite{FF64,LO64}. In contrast to the previous
studies, we have considered here strictly 1D SOC which has been realized in
ultracold Fermi gases recently~\cite{SOFermi1,SOFermi2}.

{\indent{\em Model system.}}--- We first consider quasi-1D optical lattices
aligned along the $x$ direction. The tight-binding Hamiltonian in the mean
field approximation can be written as,
\begin{equation}
H^{tb}=H_0+H_{\perp}+H_{\Delta}  \label{tbeqn}
\end{equation}
The first term $H_0=-t\sum_{\bm{i}\sigma}(c_{\bm{i},\sigma}^{\dagger}c_{%
\bm{i}+ \hat{e}_x,\sigma}+H.c.)-\mu\sum_{\bm{i},\sigma}n_{\bm{i},\sigma}+%
\frac{\alpha}{2} \sum_{\bm{i}}(c_{\bm{i}-\hat{e}_x,\downarrow}^{\dagger}c_{%
\bm{i}\uparrow}-c_{\bm{i} +\hat{e}_x,\downarrow}^{\dagger}c_{\bm{i}%
\uparrow}+H.c.) -V_z\sum_{\bm{i}}(c_{\bm{i} \uparrow}^{\dagger}c_{\bm{i}%
\uparrow}-c_{\bm{i}\downarrow}^{\dagger}c_{\bm{i}\downarrow}) -V_y\sum_{%
\bm{i}}(-ic_{\bm{i}\uparrow}^{\dagger}c_{\bm{i}\downarrow}+ic_{\bm{i}
\downarrow}^{\dagger}c_{\bm{i}\uparrow})$ is the Hamiltonian of the parallel
chains along $x$ direction where $c_{\bm{i},\sigma}^\dagger$ is the
fermionic operator creating a particle with spin $\sigma$ at site $\bm{i}={%
(i_x,i_y,i_z)}$. $t$ is the tunneling strength along $x$ direction, $\mu$ is
the chemical potential, $\alpha$ is the 1D SOC strength, $V_z$ and $V_y$ are
the out-of-plane and in-plane Zeeman fields in the $z$ and $y$ directions
respectively. All the parameters are tunable in realistic cold atom
experiments. We choose $t$ as the energy unit in this paper and set $t=1$.
The 1D SOC is realized by Raman coupling of two hyperfine states of the
cold atoms~\cite{SOBEC1,SOBEC2,SOBEC3,SOBEC4,SOFermi1,SOFermi2}. $V_z$ is
determined by the intensities of the Raman lasers and could be tuned
arbitrarily and $V_y$ could be tuned in a wide range by changing the
detuning of the lasers. 

$H_{\perp}=-t_y\sum_{\bm{i}\sigma} (c_{\bm{i},\sigma}^{\dagger}c_{\bm{i}+%
\hat{e}_{\perp},\sigma}+H.c.)$ is the spin independent transverse tunneling.
Here $\hat{e}_{\perp}=\hat{e}_{y(z)}$ is the transverse unit vector. And the
third term of the tight-binding Hamiltonian is $H_{\Delta}=\sum_{\bm{i}%
}(\Delta_i c_{\bm{i}\uparrow}^{\dagger}c_{\bm{i}\downarrow}^{\dagger} +H.c.)$,
where $\Delta_i\equiv\Delta(x_i)=-U\langle c_{\bm{i}\downarrow}c_{\bm{i}\uparrow} \rangle$.
The interaction strength $U$ between the atoms could be tuned using a Feshbach resonance,
and the Feshbach resonances in a spin-orbit coupled Fermi gases are recently
observed~\cite{FRIan}. The presence of the in-plane Zeeman field $V_y$ and
SOC breaks the spatial inversion symmetry of the Fermi surface, leading to a
FF superfluid with a finite momentum for the Cooper parings
$\Delta_i=\Delta_0e^{iQ_xx_i}$~\cite{ZZ1,ZZ2}. In an appropriate parameter regime, the system
could be driven to a topological FF superfluid phase which supports MFs~\cite%
{topoFF1,topoFF2,topoFF3,topoFF4,Chan}. In reality, the finite momentum $Q_x$ and
the order parameter $\Delta_0$ needs to be determined self-consistently~\cite{Xu}.
For simplicity, in the following analysis we enforce $Q_x=0$ and consider
the self-consistent solutions later.

To illustrate the basic physical picture of MFs in 2D and 3D optical lattices, we
first consider the quasi-1D optical lattices with a finite number $N_y$ of
chains in the transverse $y$ direction. The multi-chain Bogoliubov-de Gennes
(BdG) equation is
\begin{eqnarray}
H_{BdG}(k_x)&=&h_0(k_x)\tau_z-V_y\sigma_y-\Delta_0\sigma_y\tau_y-t_y\tau_z%
\rho_x  \label{Eqn2}
\end{eqnarray}
where $h_0(k_x)=-2t\cos{k_x}-\mu-V_z\sigma_z+\alpha\sin{k_x}\sigma_y$. The $%
4N_y{\times}4N_y$ matrices $\sigma_i$, $\tau_i$, $\rho_x$ act on the spin,
particle-hole, and chain space respectively. They are defined as $\sigma_i=%
\tilde{\rho}_0\otimes\tilde{\tau}_0\otimes\tilde{\sigma}_i$, $\tau_i=\tilde{%
\rho}_0\otimes\tilde{\tau}_i\otimes\tilde{\sigma}_0$, $\rho_x=\tilde{\rho}%
_x\otimes\tilde{\tau}_0\otimes\tilde{\sigma}_0$ where $%
\tilde{\tau}_i$ and $\tilde{\sigma}_i$ are both the $2\times 2$ Pauli
matrices, $\tilde{\rho}_x$ is a $N_y{\times}N_y$ matrix defined as $(\tilde{%
\rho}_x)_{ij}=1$ for $|i-j|=1$ and $0$ otherwise. The above BdG Hamiltonian
preserves a particle-hole symmetry $\Xi H(k_x) \Xi^{-1}=-H(-k_x)$, where $%
\Xi=\tau_x\mathcal{K}$ and $\mathcal{K}$ is the complex conjugate.

\begin{figure}[tbp]
\centering
\includegraphics[width=3.2in]{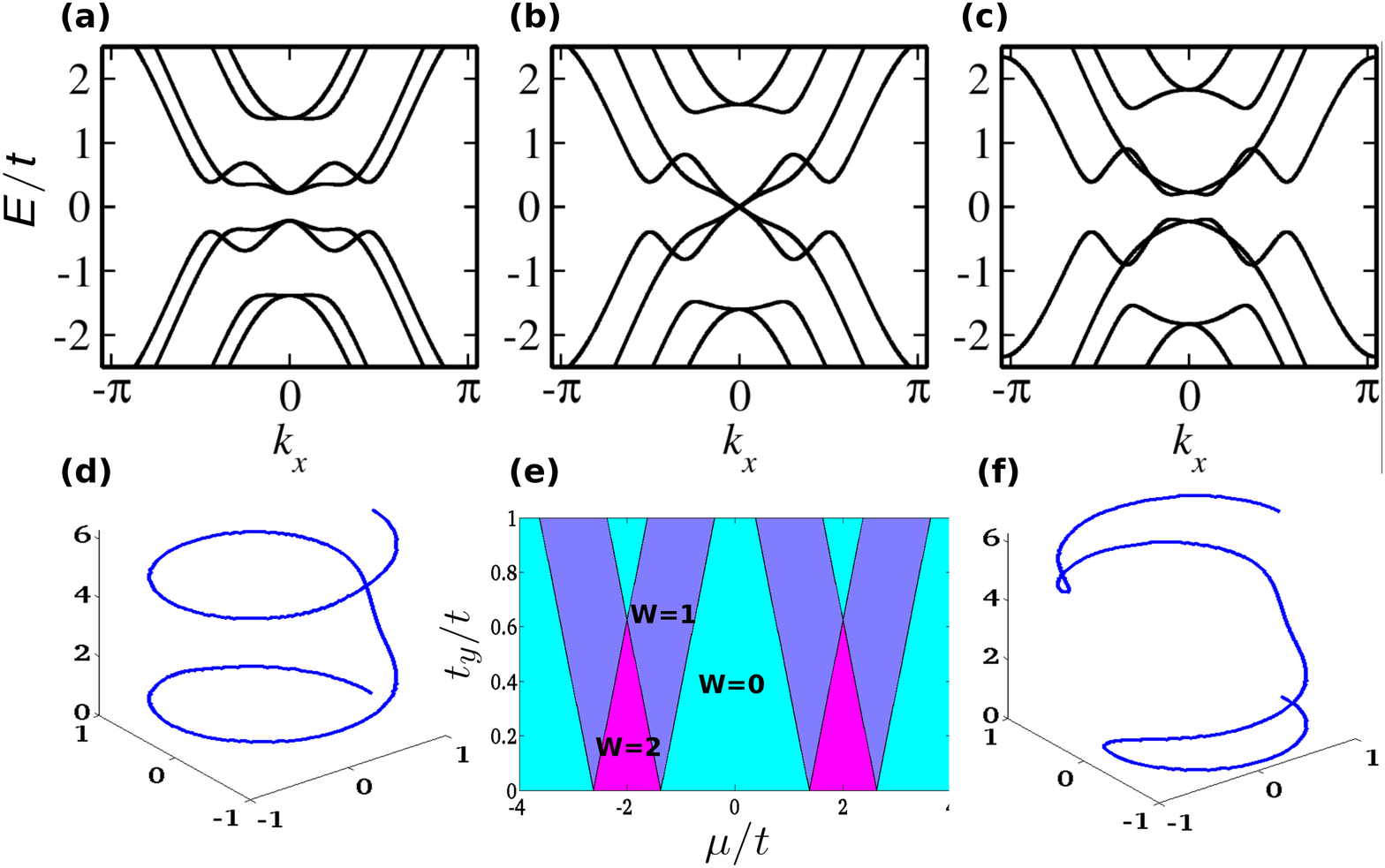}
\caption{BdG band structure of two coupled chains with increasing the
transverse tunneling for $\protect\mu=-2t$: (a) $t_y=0.3t$, (b) $%
t_y^c=0.6245t$, (c) $t_y=0.9t$. For $t_y<t_y^c$, the system is in a
topological state with the presence of two MFs, the winding number $W=2$ as
shown in (d); For $t_y>t_y^c$, the system is in a non-topological state, the
winding number $W=0$ as shown in (f). (e) is the phase-diagram of the system
indicating the number of MFs. Other parameters: $\protect\alpha=1.0t$, $%
\Delta_0=0.5t$, $V_z=0.8t$.}
\label{N2tube}
\end{figure}

{\indent{\em Chiral symmetry protected MFs for $V_y=0$.}}--- When $V_y=0$,
the above Hamiltonian preserves an additional chiral symmetry $\mathcal{S}
H(k_x)\mathcal{S}^{-1} =-H(k_x)$, where $\mathcal{S}=\tau_x$. With an
auxiliary pseudo-time reversal symmetry operator defined as $\Theta=\mathcal{%
K}$ such that $\Theta\cdot\Xi=\mathcal{S}$, the system belongs to
the BDI topological class characterized by an integer $\mathbb{Z}$
topological invariant~\cite{Class,Teo-2010}. The presence of the chiral
symmetry $\mathcal{S}$ allows the definition of a winding number $W$ which
is equal to the number of MFs in a quasi-1D system~\cite{Sumanta1,Sumanta2}.
As along as $W$ is non-zero, the multiple MFs at the ends
of the chains, even if coupled by the transverse coupling, are topologically
protected by the chiral symmetry $\mathcal{S}$. Such a chiral symmetry
cannot be broken by disorder (either site- or bond-disorder), near- or
next-near-neighbor coupling, and Zeeman fields (or magnetic impurities) in $x
$ or $z$ directions, rendering the MFs robust against all these perturbations.

As a concrete example, consider a two-chain system with weak transverse
tunneling $t_{y}$ which is the simplest quasi-1D system. The winding number
is shown in Fig.~\ref{N2tube} which agrees very well with the band structure
of the BdG Hamiltonian. We have chosen the chemical potential to lie in the
middle of the Zeeman splitting of the two uncoupled chains $\mu =-2t$. Thus,
for uncoupled chains ($t_{y}=0$) both lattices are in the topological phase
when $V_{z}>\Delta _{0}$, supporting a total of two MFs. Increasing the
transverse tunneling $t_{y}$, we see that the winding number changes from $%
W=2$ to $W=0$ at $t_{y}=t_{y}^{c}=\sqrt{V_{z}^{2}-\Delta _{0}^{2}}=0.6245t$
where the band gap closes(Fig.~\ref{N2tube}(b)). The band gap closes at $%
k_{x}=0$ at $t_{y}=t_{y}^{c}$ signalling a topological phase transition and
the disappearance of the MFs. In Fig.~\ref{N2tube}(e), we show the phase
diagram of a two-chain system indicating the number of MFs for different
parameters. The number of MFs could by any value no more than the number of
chains in the transverse direction $N_{y}$. As increasing $t_{y}$, $W$ can
change from $2\rightarrow 1\rightarrow 0$ for $|\mu |\neq {2t}$. 

\begin{figure}[tbp]
\centering
\includegraphics[width=3.2in]{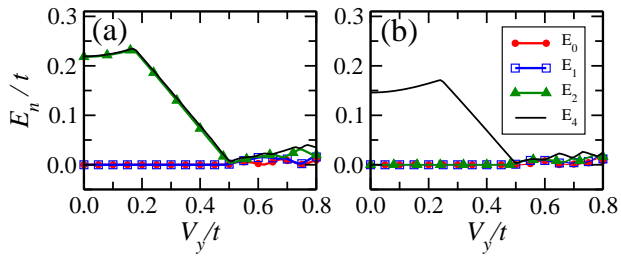}
\caption{The lowest four quasi-particle excitation energies of weakly
coupled chains for (a) $N_{y}=2$ and (b) $N_{y}=3$ as a function of the
in-plane Zeeman field $V_{y}$. The other parameters are $\protect\alpha =1.0t
$, $t_{y}=0.3t$, $\protect\mu =-2t$, $\Delta _{0}=0.5t$, $V_{z}=0.8t$.}
\label{Vychain}
\end{figure}

{\indent{\em Multiple MFs without chiral symmetry for $V_y\neq{0}$.}}--- For
a nonzero in-plane Zeeman field $V_{y}$, we see that the pseudo-time
reversal symmetry $\Theta =\mathcal{K}$ is broken and it's no longer
possible to find a chiral symmetry operator to be anti-commute with the full
BdG Hamiltonian (see Eqn.~\ref{Eqn2}). As a result, the system no longer
belongs to the BDI class~\cite{Sato-2013,Sumanta1,Sumanta2}. However, we
find that there are still multiple MFs even for $V_{y}\neq {0}$ in the
weakly coupled multi-chain system when the order parameters along different
chains are assumed to be identical. We plot the lowest four quasi-particle
excitation energies of the BdG equation for $N_{y}=2$ and $%
N_{y}=3$ in Fig.~\ref{Vychain}. We find that there are two and three zero
energy modes respectively and the system enters a gapless topologically
trivial region only for $V_{y}>\Delta _{0}$.

To understand this surprising result, we note that the multi-chain system is
of the form $H_{BdG}(k_{x})=H(V_{y})\rho _{0}-t_{y}\tau _{z}\rho _{x}$ which
can be rotated in the chain space to~\cite{Supp}
\begin{equation}
H_{BdG}(k_{x})=H(V_{y})\rho _{0}-t_{y}\tau _{z}\rho _{z}
\end{equation}%
where $\rho _{z}=\tilde{\rho}_{z}\otimes \tilde{\tau}_{0}\otimes \tilde{%
\sigma}_{0}$ and $\tilde{\rho}_{z}$ is a diagonal matrix consisting of all
the eigenvalues of $\tilde{\rho}_{x}$ (Eqn.~\ref{Eqn2}). The second term is
nothing but an effective chemical potential $\tilde{\mu}=t_{y}{(\tilde{\rho}%
_{z})_{ii}}$ for each transformed chain (they are different for different
chains). We note that the eigenvalues of $\tilde{\rho}_x$ are of the form
$(\pm \lambda_1, \pm \lambda_2, \pm \lambda_3, \cdots)$ for $N_y=even$ and
$(0, \pm \lambda_1, \pm \lambda_2, \pm \lambda_3, \cdots)$ for $N_y=odd$.
There is no essential differences between these two cases (we will show
later that the number of chains has an important effect when there is a weak
transverse SOC). After the rotation
in the chain space the new transformed ``chains" are now
independent. As long as the transverse tunneling induced effective chemical
potentials are small, the weakly coupled multi-chain system will thus have
the same number of MFs as that with $t_{y}=0$. Since the transformed chains
(each in class D with a $\mathbb{Z}_{2}$ invariant) are independent of each
other, such a system can be taken as belonging to the topological class $%
\mathbb{Z}_{2}\otimes \mathbb{Z}_{2}\cdots \otimes \mathbb{Z}_{2}=\mathbb{Z}%
_{2}^{\otimes {N_{y}}}$ for weak transverse tunneling.

{\indent{\em Effect of a transverse SOC.}}--- For comparison and
completeness, we consider a solely weak SOC between the chains in the transverse
direction for $V_y=0$, $H_{so}^{\perp }=\alpha _{y}/2\sum_{\bm{i}}(ic_{\bm{i}-\hat{e}%
_{y},\downarrow }^{\dagger }c_{\bm{i}\uparrow }-ic_{\bm{i}+\hat{e}%
_{y},\downarrow }^{\dagger }c_{\bm{i}\uparrow })+H.c$, which forms
essentially a 2D Rashba SOC but with a difference in the magnitudes between
the longitudinal and the transverse directions. We find that the effect of $%
\alpha _{y}$ is dramatically different from the effect of $t_{y}$ in the
above analysis. The chiral symmetry is no longer preserved and the system
belongs to the D topological class: for even number of chains, the system is
in the topologically trivial phase and for odd number of chains it is in the
non-trivial $\mathbb{Z}_{2}$ phase with one MF at each end. This can also be
understood from the following analysis.

The multi-chain system is of the form $H_{BdG}(k_x)=H(V_y)\rho_0-\alpha_y%
\sigma_x\rho_y$ where we have assumed that there's no transverse tunneling $t_y$.
Here $\rho_y=\tilde{\rho}_y\otimes\tilde{\tau}_0\otimes\tilde{%
\sigma}_0$, and $\tilde{\rho}_y$ is a $N_y{\times}N_y$ matrix defined as $(%
\tilde{\rho}_y)_{ij}=-i$ for $j-i=1$, $(\tilde{\rho}_y)_{ij}=i$ for $j-i=-1$
and $0$ otherwise. It has the same eigenvalues as $(\tilde{\rho}_x)$ which
form the diagonal matrix $\tilde{\rho}_z$. So the above Hamiltonian can be
rotated in the chain space to~\cite{Supp}
\begin{equation}
H_{BdG}(k_x)=H(V_y)\rho_0-\alpha_y\sigma_x\rho_z
\end{equation}
Before the rotation, the particle-hole symmetry preserves with the presence
of the transverse SOC. However, after the rotation the transverse SOC term
is now $H_{\alpha_y}=-\alpha_y(\tilde{\rho}_z)_{ii}\sigma_x$ for each
transformed chain. We find that $\Xi H_{\alpha_y} \Xi^{-1}\neq{-H_{\alpha_y}}
$, thus the particle-hole $\mathbb{Z}_2$ symmetry is broken in this new
basis and thus the original multiple MFs for $\alpha_y=0$ disappear because
of the finite transverse SOC. An exception occurs for odd number of chains
where the chain spin matrix $\tilde{\rho}_y$ has a $\lambda=0$ eigenvalue.
For this specific transformed chain, the particle-hole symmetry is intact
and thus it still has a $\mathbb{Z}_2$ symmetry which supports one MF at
each end~\cite{Sato-2013}. 

\begin{figure}[tbp]
\centering
\includegraphics[width=3.2in]{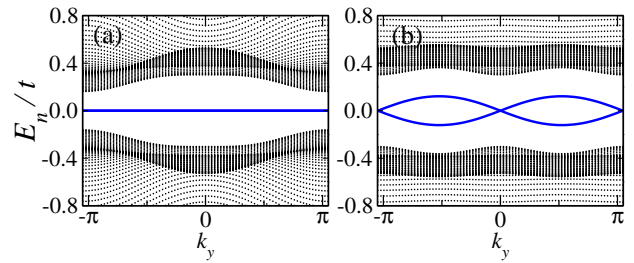}
\caption{The BdG quasi-particle excitation energies as a function of $k_y$
for a 2D strip confined in the $x$ direction with (a) a weak transverse
tunneling $t_y=0.1t$ or (b) a weak transverse SOC $\protect\alpha_y=0.2t$.
Other parameters are: $\protect\alpha=1.0t$, $V_z=0.8t$, $\Delta_0=0.5t$, $%
V_y=0.0$.}
\label{rashba2D}
\end{figure}

To compare the different effects of weak transverse tunneling $t_y$ and $\alpha_y$, we plot the edge
states of a 2D strip (confined in the $x$ direction and infinite in the $y$ direction) in these cases.
With only a weak SOC in the transverse direction $\alpha_y=0.2t$, there exist chiral MFs, with the edge
state energy spectrum plotted in Fig.~\ref{rashba2D}(b), in contrast to the energy spectrum with only a
weak transverse tunneling $t_y$ in Fig.~\ref{rashba2D}(a). We see $\alpha_y$ induces a spin dependent splitting, and its effect vanishes at $k_y=0,\pm\pi$. While for a nonzero $t_y$, the zero energies preserve the spin degeneracies and thus
form a flat band in the full parameter regime of $k_y$ as long as $t_y$ is small.

\begin{figure}[tbp]
\centering
\includegraphics[width=3.2in]{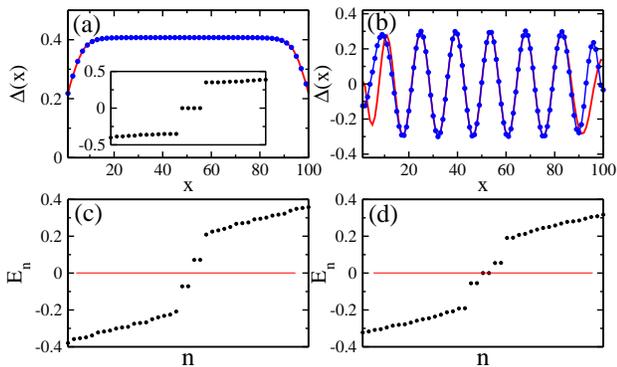}
\caption{Effects of inhomogeneous order parameters from the self-consistent
calculation. (a) The order parameters along different chains for $V_{y}=0.0$%
. The inset shows the quasi-particle excitation energies, indicating two MFs
at each end. (b,c) The order parameters (b) and the quasi-particle excitations
energies in a two-chain system with $V_{y}=0.5$. Only the real parts of the
order parameters for the two chains are plotted in (b). (d) The quasi-particle
excitation energies of a three-chain system for $V_{y}=0.5$. The order
parameter structures are similar as (b). Other parameters are $\protect%
\alpha =2.0t$, $U=4.5t$, $V_{z}=1.2t$, $\protect\mu =-2.25t$.}
\label{selfconsist}
\end{figure}

{\indent{\em Effect of inhomogeneous order parameter.}}--- In the above
discussions, we have assumed a uniform order parameter $\Delta _{0}$. In
practice, the order parameter of the superfluid needs to be obtained from the
self-consistent calculations of the BdG equation, which is naturally
inhomogeneous due to the hard wall boundary. Without an in-plane Zeeman
field $V_{y}$, the order parameters for the topological BCS superfluid are
still identical on different chains as shown in Fig.~\ref{selfconsist}(a).
The chiral symmetry guarantees the $\mathbb{Z}$ invariant and multiple MFs
are found in a quasi-1D system (here $N_{y}=2$) after a self-consistent
calculation. With a nonzero $V_{y}$, the FF superfluid order parameters of
the system are identical along different chains for periodic boundary
conditions~\cite{Xu}. However, for open boundary conditions, the order parameters are
not identical on the edges for different chains due to the interplay of the
finite $Q_{x}$ and the boundary, or the presence of edge states. As shown in
Fig.~\ref{selfconsist}(b), we see that the order parameters for the two parallel
chains are identical in the bulk but different on the boundaries. From the
quasi-particle excitation spectrum (Fig.~\ref{selfconsist}c), the multiple MFs
are gapped out because of the inhomogeneity of the order parameters on the
boundaries. However, for an odd number of chains ($N_{y}=3$), even though
the order parameters are still inhomogeneous on the boundaries, one pair of MFs
remains as shown in Fig.~\ref{selfconsist}d. More numerical results show that
one pair of MFs survives in self-consistent calculations only when $N_{y}$ and
$N_{z}$ are both odd for systems with a nonzero~$V_{y}$ \cite{Supp}.

\begin{figure}[tbp]
\centering
\includegraphics[width=3.2in]{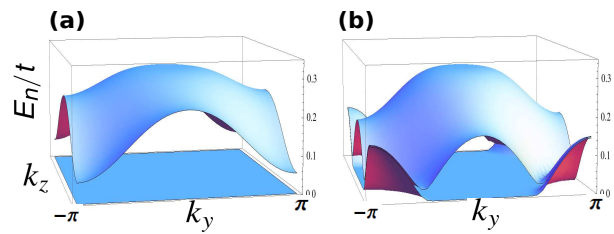}
\caption{Flat band of 3D optical lattices with weak transverse tunneling $t_y
$. (a) $t_y=0.07t$ where the Majorana zero energy states form a flat band.
(b) $t_y=0.1t$, where MFs disappear when the bulk gap at BZ zone boundaries
closes and then reopens. Only the lowest two quasi-particle excitations are
plotted to be clear. Other parameters are: $V_z=0.8t$, $\Delta_0=0.5t$, $%
\protect\alpha=1.0t$, $V_y=0.0$. The chemical potential $\protect\mu=-2t-4t_y
$ which removes the transverse tunneling induced constant energy offset.}
\label{flatband}
\end{figure}

{\indent{\em Majorana flat band.}}--- We now study the crossover of the
Majorana fermion physics from quasi-1D to a real 3D system. Assume that
the 3D lattices with strictly 1D SOC
are confined in the $x$ direction. With periodic boundary conditions along
the transverse directions, $k_{y}$ and $k_{z}$ are good quantum numbers and
serve as external parameters. Such 3D lattices are equivalent to a quasi-1D
system with a number of chains equal to the discrete values of $k_{y/z}$.
The chemical potential is assumed to occupy the middle of the lower bands
when the transverse tunneling $t_{y}$ is zero. We also take a Zeeman field
large enough to drive the system to the topological regime for each value of
$k_{y}$ and $k_{z}$. Thus, the lowest quasi-particle excitations of the
uncoupled 3D lattices shows a MFs flat band~\cite{Volovik,flatband}. The
zero energy MFs flat band persists for a small transverse tunneling $t_{y}$
(Fig.~\ref{flatband}a). With increasing $t_{y}$,
the zero energy MFs disappear first at the Brillouin zone (BZ) edges while
the BZ center is still a zero energy flat band (Fig.~\ref{flatband}b).
Further increasing $t_{y}$, we see that the MFs flat band is enclosed by the
bulk quasi-particle excitations, showing a configuration of Majorana flat
plate. When the transverse coupling is strong enough, MFs disappear from the
entire BZ, transforming the system to a trivial superconductor. Note that
the bulk excitation gap at the center of BZ also decreases. These
demonstrate that MFs exist at the confined surface of a 3D system and the
zero energy flat band is topologically protected by an energy gap from the
bulk excitations.

{\indent{\em Conclusion.}}--- In summary, we have studied MFs in quasi-1D, 2D, and
3D optical lattices in the presence of weak transverse tunnelings between the chains. The
realistic experimentally realized SOC is strictly 1D, leading to new intriguing
results comparing to systems with a 2D Rashba SOC. We find that,
topologically robust MFs exist even in a coupled multi-chain system
as long as the transverse couplings are weak. We have studied
the robustness of the MFs to in-plane Zeeman fields (that break the chiral symmetry)
and also to spatially inhomogeneous order parameters resulting from the
self-consistent calculations. For the 3D optical lattices confined in the SOC
direction, a zero energy MFs flat band may exist even in the presence of a weak
transverse tunneling. The existence of MFs in cold atom systems can be
probed using radio-frequency spectroscopy~\cite{RF1,RF2}. Spin-orbit coupled
fermionic optical lattices has been experimentally realized. Because
of the lack of disorder and the precisely controllable experiment parameters,
it provides a feasible platform to search for MFs in context of cold atoms.

\textit{Acknowledgement:} C.Q., Y.X. and C.Z. are supported by ARO
(W911NF-12-1-0334), AFOSR (FA9550-13-1-0045), and NSF-PHY (1104546). M.G. is
supported by Hong Kong RGC/GRF Projects (No. 401011 and No. 2130352)
and the Chinese University of Hong Kong (CUHK) Focused Investments Scheme.
S.T. is supported by NSF (PHY-1104527) and AFOSR (FA9550-13-1-0045).
S.T. would like to thank Tudor \ref{•}Stanescu for discussions.

\begin{widetext}
\newpage
\section{supplementary material}
\subsection{Rotation in the chain space for $t_y{\neq}0$ and $\alpha_y=0$}
The BdG Hamiltonian is of the form
\begin{equation}
H_{BdG}(k_x)=H\rho_0-h_x\rho_x
\end{equation}
where $H=[-2t\cos{k_x}-\mu-V_z\sigma_z+\alpha\sin{k_x}\sigma_y]\tau_z-\Delta_0\sigma_y\tau_y
-V_y\sigma_y$ and $h_x=t_y\tau_z$. A unitary transformation
$U=\tilde{U}\otimes\tilde{\sigma}_0\otimes\tilde{\sigma}_0$ in the chain space gives
\begin{equation}
H_{BdG}=H\rho_0-h_x\rho_z
\end{equation}
where $\rho_z=\tilde{\rho}_z\otimes\tilde{\sigma}_0\otimes\tilde{\sigma}_0$ and $\tilde{U}$ is a $N_y{\times}N_y$ matrix which diagonalizes
$\tilde{\rho}_x$ as $\tilde{U}\tilde{\rho}_x\tilde{U}^{-1}=\tilde{\rho}_z$.
Thus, for each transformed chain the transverse tunneling term is now
$-t_y({\tilde{\rho}_z})_{ii}\tilde{\tau}_z$
, which is an effective
chemical potential. Explicitly,
for a system with a number of chains $N_y$ in the $y$ direction
\begin{equation}
H_{BdG}=
\left[
\begin{array}{cccc}
H-t_y({\tilde{\rho}}_z)_{11}\tilde{\tau}_z  &  0  &  \vdots  &  0 \nonumber \\
0 & H-t_y(\tilde{\rho}_z)_{22}\tilde{\tau}_z & \vdots & 0 \nonumber \\
\ldots & \cdots & \ddots & \cdots \nonumber \\
0 & 0 & \vdots & H-t_y(\tilde{\rho}_z)_{N_yN_y}\tilde{\tau}_z
\end{array}
\right]
\end{equation}
As long as the transverse tunneling induced effective chemical potentials are small, there would
be no quantum phase transition, and the number of MFs is equal to that without the transverse tunneling.

\subsection{Rotation in the chain space for $t_y=0$ and $\alpha_y{\neq}0$}
The BdG Hamiltonian is of the form
\begin{equation}
H_{BdG}(k_x)=H\rho_0-h_y\rho_y
\end{equation}
where $h_y=\alpha_y\sigma_x$. A similar unitary transformation in the chain space gives
\begin{equation}
H_{BdG}=H\rho_0-h_y\rho_z
\end{equation}
where $\rho_z=\tilde{\rho}_z\otimes\tilde{\sigma}_0\otimes\tilde{\sigma}_0$ and its reduced matrix $\tilde{\rho}_z$ is a diagonal matrix whose elements are the eigenvalues of $\tilde{\rho}_y$.
For each transformed chain, such a term is of the form $-\alpha_y(\tilde{\rho}_z)_{ii}\tilde{\sigma}_x$
which breaks the particle-hole symmetry operator in the new basis. An exception occurs for an odd number
of chains because one of the eigenvalues is zero, i.e $(\tilde{\rho}_z)_{\frac{N_y}{2}\frac{N_y}{2}}=0$ if the diagonal elements
are arranged in an ascending order.
Explicitly, for a system with a number of chains $N_y$ in the $y$ direction
\begin{equation}
H_{BdG}=
\left[
\begin{array}{cccc}
H-\alpha_y(\tilde{\rho}_z)_{11}\tilde{\sigma}_x  &  0  &  \vdots  &  0 \nonumber \\
0 & H-\alpha_y(\tilde{\rho}_z)_{22}\tilde{\sigma}_x & \vdots & 0 \nonumber \\
\ldots & \cdots & \ddots & \cdots \nonumber \\
0 & 0 & \vdots & H-\alpha_y(\tilde{\rho}_z)_{N_yN_y}\tilde{\sigma}_x
\end{array}
\right]
\end{equation}

\begin{figure}[tbp]
\centering
\includegraphics[width=3.5in]{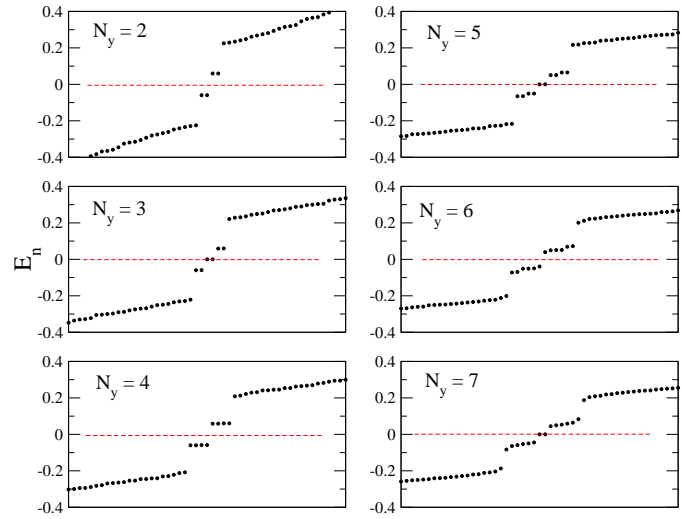}
\caption{The quasi-particle excitation energies for a quasi-1D system with
a number of $N_y\times 100$ chains after a self-consistent calculation. $N_z=1$.}
\label{supp1}
\end{figure}

\subsection{More self-consistent solutions for quasi-1D topological FF superfluid}
Fig.\ref{supp1} and Fig.\ref{supp2} are more examples of the multi-chain topological
FF superfluid after a self-consistent calculation. We see that there's no zero energy states
when the number of chains along any transverse direction is even and only one pair of MFs persist when $N_y$ and $N_z$ are both odd numbers.

\begin{figure}[tbp]
\centering
\includegraphics[width=3.5in]{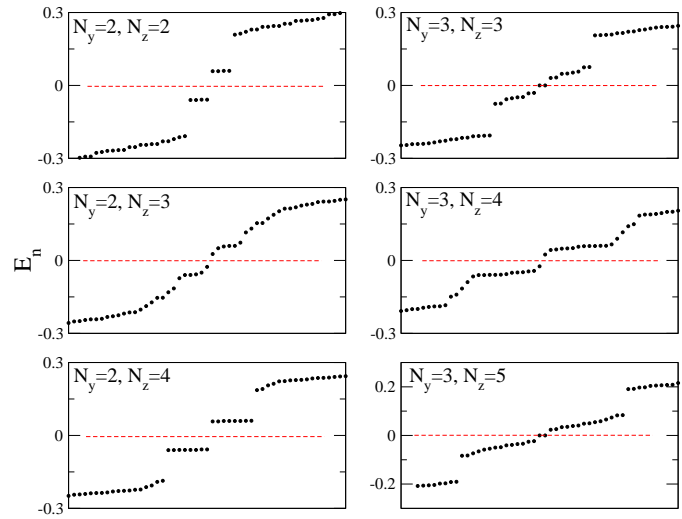}
\caption{The quasi-particle excitation energies for a quasi-1D system with
a number of $N_y\times N_z\times 100$ chains after a self-consistent calculation.}
\label{supp2}
\end{figure}

\end{widetext}
\end{document}